\documentstyle[aps,prb,preprint,floats]{revtex}
\input psfig.tex

\def\wt{\tilde}
\def\ltwid{\mathrel{\raise.3ex\hbox{$<$\kern-.75em\lower1ex\hbox{$\sim$}}}}
\def\gtwid{\mathrel{\raise.3ex\hbox{$>$\kern-.75em\lower1ex\hbox{$\sim$}}}}
\def\YBCO{$YBa_2 Cu_3 O_{7-\delta}$}
\def\dxy{d_{x^2-y^2}}
\def\im{{\rm Im}}

\begin{document}
\vskip -2cm
\title{$\pmb{}$ \pmb{$c$}-Axis Infrared Conductivity of a \pmb{$\dxy$}-Wave
Superconductor with Impurity and Spin-Fluctuation Scattering}
\author{P. J. Hirschfeld}
\address{Department of Physics,
University of Florida,
Gainesville, FL 32611}
\author{S. M. Quinlan }
\address{Solid State Division, Oak Ridge National Laboratory,
P. O. Box 2008, Oak Ridge, TN 37831--6032\\
and\\
Department of Physics and Astronomy,
University of Tennessee
Knoxville, TN 37996--1200}
\author{D. J. Scalapino }
\address{Department of Physics, University of California,
Santa Barbara, CA 93106--9530}
\date{August 3, 1996}
\maketitle
\vskip -1cm
\begin{abstract}
Results of a calculation of the $c$-axis infrared conductivity $\sigma_c$
for a $\dxy$-wave superconductor which include both elastic impurity and
inelastic spin-fluctuation scattering are presented
and compared with the $ab$-plane conductivity $\sigma_{ab}$ in the same model.
In this model, the interlayer $c$-axis coupling is taken to be
weak and diffusive.
While in clean systems, inelastic scattering leads to a peak
at $\omega=4\Delta_0$ in $\sigma_{ab}$ for $T<T_c$, it has
little effect on the corresponding $\sigma_c$, which exhibits structure
only at $\omega\simeq 2\Delta_0$ and is directly related to the
single-particle density of states $N(\omega)$.  The $c$-axis penetration
depth $\lambda_c$ in the same model is predicted to vary as $T^3$
at low temperatures in clean samples.  We discuss recent optical
experiments on the cuprates and compare with these predictions.
\vskip .0cm
\noindent PACS: 74.25.Nf, 74.72.-h, 72.10.Di, 72.10.Fk
\end{abstract}

\section{Introduction}
In the superconducting state of optimally doped \YBCO, a gap-like
depression develops in the frequency dependence of the $c$-axis
conductivity for $T$ less than $T_c$.
As emphasized by Homes {\it et al.},\cite{Homes,Basov} this behavior is quite
different from the in-plane $ab$ conductivity observed in samples from
the same source.
We have recently developed a simple model for the $ab$ conductivity
which provided a qualitative fit to experimental data, and here we
extend this model to see how well it can describe the $c$-axis
conductivity.
The model consists of BCS quasi-particles with a $\dxy$ gap and a
lifetime determined by both elastic and inelastic scattering processes.
The latter inelastic lifetime plays a key role in our model and was
calculated from a spin-fluctuation interaction in which the
quasi-particle $\dxy$ gap entered explicitly in determining the spectral
weight of the spin-fluctuations.
The elastic scattering, which is important at low frequencies, was
treated in the resonant impurity scattering limit.
Here, in discussing the $c$-axis conductivity, we will treat the
interlayer transport as weak and diffusive.
Our conclusion from this analysis is that a model of a $\dxy$
superconductor which includes inelastic as well as elastic scattering
can provide a sensible fit of both the $ab$ and the $c$-axis
conductivities.  In particular, it predicts the existence of absorption
features at $4\Delta_0$ in $\sigma_{ab}$ and at $2\Delta_0$ in
$\sigma_c$, where $\Delta_0$ is the $d$-wave gap maximum; such features
are apparently observed in a variety of materials.
To the extent that the interlayer coupling can be treated as
weak and diffusive,
the $c$-axis conductivity provides direct
information on the single particle density of states in a similar manner
to SIS tunneling.
\section{\protect\bbox{c}-axis conductivity}
The nature of $c$-axis transport in the cuprates is still controversial,
but it is widely believed to be weak and incoherent.
\cite{Leggett,Grafetal,RojoLevin,Radtke}
A simple model which describes this
has $H=H_0+H_c$, where $H_0$ describes free
particles with residual $d_{x^2-y^2}$ BCS pair interaction, and
\begin{equation}
H_c=\sum_{i\sigma}V_i (c_{i1\sigma}^\dagger c_{i2\sigma} + h.c.),
\end{equation}
where the $V_i$ may be thought of as randomly placed static impurities
mediating
hopping between planes 1 and 2.  We emphasize, however, that the end
result of our treatment will yield a description in terms of a single
$c$-axis relaxation time which must be similar to any weak diffusive
transport model.  The interplane paramagnetic current operator for a given
configuration of the random potential is then
\begin{equation}
j_c^p = -ied\sum_i V_i (c_{i1\sigma}^\dagger c_{i2\sigma} - h.c.).
\end{equation}
The response to an applied electromagnetic field ${\bf A}$
is now determined by expanding
the kinetic energy  of the system up to second order in ${\bf A}$,
yielding for the expectation value of the
total current
\begin{equation}
\langle j_c ({\bf q},\omega)\rangle = [d^2 e^2 \langle H_c \rangle
+ \Lambda_{cc} ({\bf q},\omega) ] A_z ({\bf q}, \omega),
\end{equation}
where $\langle j_c^p \rangle_{\bf A} \equiv \Lambda_{cc}A_z$ to leading
order.

We first examine the conductivity $\sigma_c (q=0,\omega)$, assuming
that $V$ is weak, thereby neglecting multiple scattering
vertex corrections as well as self-energy corrections, since one
power of $V$ is already contained in each of the $c$-axis current
operators which enter $\Lambda_{cc}$.  After impurity averaging,
we find
\begin{equation}
\sigma_c(\omega) = - {\im \Lambda_{cc}(\omega)\over\omega} ,
\end{equation}
with
\begin{eqnarray}
\im\,\Lambda_{cc}(\omega) =~~~~~~~~~~~~~~~~\nonumber\\
4\,\pi\,d^2e^2 \int d\omega'
\sum_{{\bf p p^\prime}}
\left|
V_{{\bf p}- {\bf p}^\prime}
\right|^2 \left( f(\omega+\omega')-f(\omega')\right)\\
\times
\left(A(p,\omega+\omega')A(p',\omega') +
B(p,\omega+\omega')B(p',\omega') \right) .\nonumber
\end{eqnarray}
Here the normal and anomalous spectral functions are defined as
$A(p,\omega) = -{1\over 2\pi} \im  {\rm Tr}\{ {\underline G}({\bf k},
\omega+i0)\} $, $B(p,\omega) = -{1\over 2\pi} \im  {\rm Tr} \{\tau_1
{\underline G}({\bf k},
\omega+i0)\}$, where ${\underline G}$ is the matrix Green's function
\begin{equation}
{\underline G}({\bf k},
\omega )= {\wt\omega\tau_0 +\xi_p\tau_3+\Delta_0\tau_1
\over\wt\omega^2-\xi_p^2-\Delta^2_p}.
\end{equation}
The interlayer scattering matrix element  $V_{{\bf p}- {\bf p}^\prime}$
is defined by
\begin{equation}
V_i = \sum_m {\cal V} ({\bf R}_i - {\bf R}_m) =
\sum_{{\bf p},m} V_p e^{ip\cdot({\bf R}_i - {\bf R}_m)},
\end{equation}
where m runs over all lattice sites and i labels an impurity site.
For simplicity we take a cylindrical Fermi surface with $\xi_p =
(p^2/2\,m-\mu)$ and a $\dxy$ gap $\Delta_p = \Delta_0(T) \cos\,2\,\phi$.
The renormalized frequency $\wt\omega$ includes the effects of  scattering
associated with the in-plane impurities and the lifetime
effects due to dynamic spin fluctuations.
In the resonant limit, the in-plane impurity scattering gives a
renormalized quasi-particle energy $\wt\omega$
determined by the self-consistent solution of
\begin{equation}
\wt\omega = \omega-\Sigma_{imp}(\wt\omega)=\omega + i\Gamma /
\left\langle {\wt\omega\over
\sqrt{\wt\omega^2-\Delta^2_k}}\right\rangle_{\rm FS}.
\end{equation}
Here $\Gamma = n_i/\pi N_0$, with $n_i$ the impurity concentration and
$N_0$ the Fermi level density of states.

We assume further that the inelastic scattering rate $\tau^{-1}_{\rm
in}(\omega,T)$ arising from  spin fluctuations can be calculated
separately, and take as a model  the
one-antiparamagnon self-energy calculated in a 2D Hubbard model
by Quinlan {\it et al.}\cite{Quinlanetal}  The momentum dependence of
the full $1/\tau_{in} ({\bf p},\omega) $ has been neglected here in favor
of a simpler effective relaxation rate for quasiparticles at the
gap nodes, $1/\tau_{in}(\omega) \equiv 1/\tau_{in}({\bf k}_{node},\omega)$
as discussed in Refs.~\onlinecite{Quinlanetal} and \onlinecite{QHS}
At low temperatures $\tau^{-1}_{\rm in}$ varies as $\omega^3$ for
$\omega\ltwid 3\,\Delta_0$, reflecting the usual $\omega^2$ dependence
of an electron-electron scattering process along with a third factor of
$\omega$ arising from the $\dxy$ density of states.
For $\omega\gtwid 3\,\Delta_0$, $1/\tau_{in}$ crosses over to the pseudo-linear
behavior found in spin-fluctuation scattering calculations in the normal
state.\cite{Moriya,MFL,Coffey2,MonPines}
Although for concreteness we have chosen
a particular model for $1/\tau_{in}$, we note that the behavior in
the superconducting state is characteristic of any model based
on electron-electron scattering with  quasiparticles which are well-defined
at sufficiently low temperatures, in the presence of gap line nodes.

Combining the elastic and inelastic contributions, we set
\begin{equation}
\wt\omega = \omega - \Sigma_{imp}(\wt\omega) + {i \over \tau_{\rm
in}(\omega,T)}
\end{equation}
in Eqs.~(8) and (9).   The resulting lifetime for in-plane scattering
is plotted in Fig.~1 for several values of the impurity concentration.
Note the Hubbard parameters have been chosen so as to give a value
of $1/\tau_{\rm in}(\omega\rightarrow 0,T_c)=2 T_c$.

As discussed by Graf {\it et al.},\cite{Grafetal}
setting $|V_{{\bf p}- {\bf p}^\prime}|^2$ equal to a
constant corresponds to diffuse transmission, while taking an extreme
forward scattering form such that $\hat p = \hat p'$ (that is,
$\phi=\phi'$) corresponds to specular transmission.
In the diffuse transmission limit, the second term in Eq.~(5), arising
from pair transfers, vanishes for a $d$-wave gap.
In the actual system, $|V_{pp'}|^2$ is anisotropic, and here we model
this using the separable form
\begin{equation}
|V_{pp'}|^2 = |V|^2 + |V'|^2 \cos\,2\,\phi\,\cos\,2\,\phi'.
\end{equation}
The interaction strength $V$ can be adjusted to fit the normal state
$c$-axis conductivity $\sigma_{cn} = {2\,d^2e^2N_0 / \tau_c}$ with
$\tau_c^{-1} = 2 n_i^c \,\pi N_0|V|^2$.  Here $n_i^c\sim {\cal O} (1)$
is formally the density of
scattering sites which allows the planes to communicate.
The second term $V'$ can be used to fit the $c$-axis penetration depth
$\lambda_c(0)$ (see below)
\begin{equation}
{c\over 4\,\pi\lambda^2_c(0)}
= 0.480 \sigma_{cn} \left|{V'\over V}\right|^2
\Delta_0(0).
\end{equation}
For $\sigma_n \sim 150\,(\Omega\,$cm$)^{-1}$ and $\lambda_c(0) \sim
10^4\,$A, we find $|V'/V|^2 = {\cal O}(1)$.

With the form (10) for $|V_{{\bf p}- {\bf p}^\prime}|^2$,
the expression Eq.~(4) for
$\sigma_c(\omega)$ simplifies to
\begin{eqnarray}
{\sigma_c(\omega)\over\sigma_{cn}}& =&
{1\over \omega}
\int d\omega' \, \left(f(\omega') - f(\omega'+\omega)\right)\\
&\times&\left(
N(\omega+\omega') N(\omega') + \left| {V' \over  V}\right|^2
M(\omega+\omega')M(\omega')\right), \nonumber
\end{eqnarray}
with
\begin{equation}
N(\omega) = {\rm Re}
\left\langle {\wt\omega \over \sqrt{\wt\omega^2-\Delta^2_p}}
\right\rangle_{\rm FS} = {2\over \pi}{\rm Re} ~
{\bf K} (\Delta_0/\wt\omega)
\end{equation}
and
\begin{eqnarray}
M(\omega) &=&
{\rm Re}\left\langle {\cos\,2\phi~ \Delta_p \over
\sqrt{\wt\omega^2-\Delta^2_p}} \right\rangle_{\rm FS}\\&=&
{2\over \pi}{\rm Re}\Big\{
{\wt\omega\over \Delta_0}[{\bf K} (\Delta_0/ \wt\omega)-
{\bf E} (\Delta_0/ \wt\omega )]\Big\}.\nonumber
\end{eqnarray}
\noindent Here $\bf K$ and $\bf E$ are the complete elliptic integrals
of the first and second kinds, respectively.\cite{GR}
Results for $\sigma_c(\omega)/\sigma_{cn}$ versus $\omega$ at different
temperatures are shown in Fig.~2
under the assumption of isotropic diffusive scattering ($|V'/ V|^2=0$).

The primary
feature seen in the figure is the pseudogap of width $2\Delta_0$ which
opens up below $T_c$.   The low-frequency behavior can be estimated
in this limit to be $\sigma_c(\omega) /\sigma_{cn}\simeq
2\omega^2/3\Delta_0^2$.  Figure 3 illustrates the effects
of adding anisotropy to the interplane scattering, as well as
intraplane scattering effects.  Because of
 the form of Eq.~(12) as a convolution
of normal and anomalous densities of states, the shoulder feature at
$2\Delta_0(T)$ is  quite robust.

While strongly scattering in-plane impurities ($\Gamma > 0$) can alter the very
low-frequency
behavior, the $2\Delta_0$ shoulder is essentially unaffected.
The large intraplane relaxation rate near $T_c$  might be expected
to change the behavior in an important way, but this is seen not
to be the case, since the effect of a large $1/\tau$ is simply to smear
out the density of states factors entering Eq.~(12).
Scattering anisotropy $(|V'/ V|^2>0)$ also does not
lead to important qualitative differences
from the results of  Fig.~2, since the anomalous terms in the
convolution give rise to a low-energy contribution to the conductivity
of  order
$(\omega/\Delta_0)^4$.  The results in these limits are therefore
similar to those obtained earlier by Graf et
al.\cite{Grafetal}  In Fig.~4 we compare these results to
the experimental data of Homes {\it et al.}; here we have chosen a parameter
set which appears to give a reasonable fit to the $ab$
penetration depth and conductivity, but the $c$-axis conductivity
is not terribly sensitive to this choice.

Due to the weak behavior of the anomalous density of states at low
temperatures,
we note that within this model a measurement of the $c$-axis conductivity
provides
a nearly direct measurement of the in-plane density of states $N(\omega)$.  It
will be interesting to compare $N(\omega)$ obtained in this way to results
obtained, e.g., from tunneling measurements.
\section{\protect\bbox{c}-axis penetration depth}
The real part of the current-current correlation function may
also be calculated, yielding
\begin{eqnarray}
\Lambda_{cc}(q\rightarrow 0, \omega=0)&=&
-2 e^2 d^2 n_i^c T \sum_{\omega_n}\sum_{pp^\prime \sigma}
|V_{{\bf p}- {\bf p}^\prime}|^2 \\&\times&{1\over 2} {\rm Tr}
[\tau_0 {\underline G}_\sigma({\bf p},\omega_n ) \tau_0 {\underline G}_\sigma
({\bf p}^\prime,\omega_n )
].\nonumber
\end{eqnarray}
The diamagnetic term in Eq.~(3) takes
the  form
\begin{eqnarray}
\langle H_c \rangle &=& 2 n_i T \sum_{\omega_n}\sum_{pp^\prime \sigma}
|V_{{\bf p}- {\bf p}^\prime}|^2
\\&\times& {1\over 2} {\rm Tr}
[\tau_3 {\underline G}_\sigma({\bf p},\omega_n) \tau_3 {\underline G}_\sigma
({\bf p}^\prime,\omega_n )
],\nonumber
\end{eqnarray}
which means that only the product of the anomalous Green's functions
survives in this approximation.

The remaining frequency sums
and integrals may be evaluated in the diffusive limit
treated above to find the London penetration depth, $\lambda_c(T)$, as
\begin{eqnarray}
{c\over 4\pi\lambda_c(T)^2} &=& - \sigma_{cn} |V'/ V|^2
\int d\omega ~{\rm \tanh} ({\omega \over 2T}) ~ {\rm Im}
[ \Phi(\omega)^2 ]\\ &\simeq &
 \sigma_{cn} |V'/ V|^2
\Delta_0(T)[0.480 - {6\zeta (3)\over \pi}\Big({T\over \Delta_0(T)}\Big)^3],
\end{eqnarray}
where $\Phi=
{2\over \pi}{\wt\omega\over \Delta_0}[{\bf K} (\Delta_0/ \wt\omega)-
{\bf E} (\Delta_0/ \wt\omega )]
$ is the same angular average appearing in Eq.~(14).
Note the last approximate equality
is valid for $T\ll T_c$ in the clean limit only.
The expression (17) is equivalent to the result
obtained by Radtke {\it et al.}\cite{Radtke} in the absence of
direct interlayer tunneling.  In Fig.~5 we have plotted
the temperature dependence of $\lambda_c (0)^2/\lambda_c(T)^2$ for various
values of the in-plane impurity scattering rate.  As in the $ab$ case,
\cite{pendepthcrossover} we expect
a crossover to a $\delta\lambda_c\sim T^2$ behavior at low temperatures
when the system becomes sufficiently dirty.  Although the normalized superfluid
density
$(\lambda_c(0)/\lambda_c(T))^2$  is quite insensitive to the presence of
planar impurities, as seen in Fig.~5, we note the peculiar result
that in this model the penetration
depth itself scales roughly with the $c$-axis conductivity, and should therefore
be
quite sensitive to disorder which affects $c$-axis transport.

\section{Comparison of \protect\bbox{ab}
and \protect\bbox{c}-axis data}
In constructing phenomenological theories of this kind, it is important to see
if a consistent description of a wide class of experiments can be obtained,
along with an understanding of the effects of disorder.  In this regard we
would
like to predict correlations among various quantities which should hold
at least for all optimally doped materials, and possibly in the underdoped
systems as well.  The most important such correlation to emerge from the
current
analysis is the prediction of a clear shoulder in the $c$-axis conductivity at
$\omega=2\Delta_0$, a broad maximum in
the $ab$ conductivity at $4\Delta_0$, and a shoulder-like feature in
$1/\tau_{ab} (\omega)$ at $3\Delta_0$.  Since
the gap maximum $\Delta_0$ is not
known for all materials,\cite{photo}
one can search for correlations of such features in optical experiments.
While such a method will not provide the most accurate measure of the gap
size, it will serve to confirm the general picture of a $d$-wave order
parameter
with nodal quasiparticles scattered by resonant defects of some kind at low
energies, and strong spin-fluctuation scattering at higher energies.  In
Fig.~6, we show $\sigma_c$, $\sigma_{ab}$, and $1/\tau_{ab}$ together to
illustrate
the correlations we expect.
In Fig.~7, we show data on $YBa_2 Cu_3 O_{6.95}$, $YBa_2 Cu_3 O_{6.6}$,
and $YBa_2Cu_4 O_8$ from
recent work by Basov {\it et al.}\cite{Basovpreprint}
Here we would like to note some similarities between the results for the
optimally doped material below its $T_c$, and the behavior of the underdoped
materials above their transition temperatures in the so-called
pseudogap regime.
The most distinct
feature is the shoulder in $\sigma_c$ (upper panel of Fig.~7),
which, as we have discussed, occurs at
$2\Delta_0\simeq 6-8 T_c$ for $YBa_2 Cu_3 O_{6.95}$.
In the underdoped compounds this shoulder, at what we will call
``$2\Delta_0$'',
becomes visible above $T_c$ in the pseudogap regime.  In the underdoped
compounds, ``$2\Delta_0$'' from this criterion is slightly reduced from its
value in the optimally doped
materials.  The second panel in Fig.~7 shows
$\sigma_{ab} (\omega)$.  In Ref.~\onlinecite{QHS},
the absence of a $2\Delta_0$ structure and the broad rise at $4\Delta_0$
observed in
the $ab$-plane conductivity of
$YBa_2 Cu_3 O_{6.95}$
below $T_c$ at 1000 ${\rm cm}^{-1}$ was attributed
to a $4\Delta_0$
inelastic absorption threshold smeared by $d$-wave nodal quasiparticles.
Now,
as is well known, there is no gaplike structure in the $ab$
conductivity (second panel of Fig.~7) in the doping range for which
the pseudogap is
observed in the $c$-axis conductivity; there is, however,
a narrowing of the Drude peak and
a broad rise at higher frequencies near ``$4\Delta_0$''.  It could
be that the appearance
of the ``$2 \Delta_0$'' in $\sigma_c$ but not
in $\sigma_{ab}$ is similar to the behavior we have found for these
quantities below $T_c$ in the model we have applied to $YBa_2 Cu_3 O_{6.95}$,
in which
the crucial factor is the direct relevance of the density of
states $N(w)$ to $c$-axis transport.
Support for the correlation of the $c$-axis pseudogap and the density of
state is found in studies by Homes {\it et al.}\cite{Homes} which found
scaling of the pseudogap with the Knight shift in underdoped cuprates.
Finally, as shown in the bottom panel of Fig.~7, that the in-plane
relaxation rate for the underdoped materials shows a strong supression
of the linear frequency dependence in the pseudogap regime even above $T_c$.
In all cases shown, the crossover to stronger frequency dependence
at sufficiently low temperatures occurs at an energy consistent with
the ``$3\Delta_0$'' smeared threshold expected from the current model.

We turn finally to the comparison of the $c$-axis penetration depth with its
$ab$ counterpart.  The results are
shown in Fig.~8 for the clean limit, and
compared to the data of Hardy {\it et al.}\cite{Hardyetal} on $YBa_2 Cu_3
O_{6.95}$.
It is clear that the result agrees qualitatively with the much flatter
temperature dependence of the $c$-axis penetration
depth found at low temperatures.  The low temperature
behavior is not yet well characterized, however.
The discrepancies result, in the intermediate temperature
range, from the crude isotropic band structure assumed, and in the
range close to $T_c$ from critical fluctuations not accounted for in the
current model.

\section{Conclusions}
We have argued that a simple theory of $d$-wave superconductivity, together
with impurity and spin fluctuation scattering, accounts for most of the
qualitative aspects of both  optical conductivity and London penetration depth
measurements for currents along both the $a$ and $c$ axes.
$c$-axis properties
have been calculated under the further assumption that  transport
is primarily diffusive.  The optical conductivities
$\sigma_c$ and $\sigma_{ab}$ may then be correlated in
this model, in that we expect
a broad peak in $\sigma_{ab}$ at $\omega\simeq 4\Delta_0$ and a shoulder-like
feature in $\sigma_c$ at $\omega \simeq 2\Delta_0$.  This behavior indeed
appears
to be reproduced qualitatively in several materials, including
underdoped materials, where a behavior qualitatively
similar
to that observed in the optimally doped materials below $T_c$ is seen
significantly above $T_c$.
If the phase coherence implicit in the BCS approach is ignored,
the model considered here offers a natural interpretation of these results
and an explanation of why the normal state $c$-axis pseudogap does not
manifest itself in planar transport properties.  Further study is needed
to put these speculations on a firmer basis.

Finally, we have calculated and compared the $c$-axis London penetration
depth in the same model, and shown that in clean systems it will vary
as $\delta \lambda_c\sim T^3$ at low temperatures.  This temperature
dependence is much stronger than  the $ab$ result, $\delta\lambda_{ab}
\sim T$ in the same model, in qualitative
agreement with experiments on optimally
doped  \YBCO.

\section{Acknowledgements}
We are grateful to C. Homes, T. Timusk, and D. Basov for providing
their data prior to publication, and for enlightening discussions.  We also
acknowledge useful discussions with W. N. Hardy and D. B. Tanner.
This work was supported by the University of Tennessee and the
Division of Materials Sciences, U. S. Department of Energy, under
Contract No.~DE--AC05--96OR22464 with Lockheed Martin Energy Research
Corp.\ (SMQ),
by NSF under grants  DMR95--27304 (DJS) and  DMR-96--00105 (PJH).
Some of the numerical calculations
reported in this paper were performed at the San Diego Supercomputer
Center.

\begin{figure}[p]
\leavevmode\centering\psfig{file=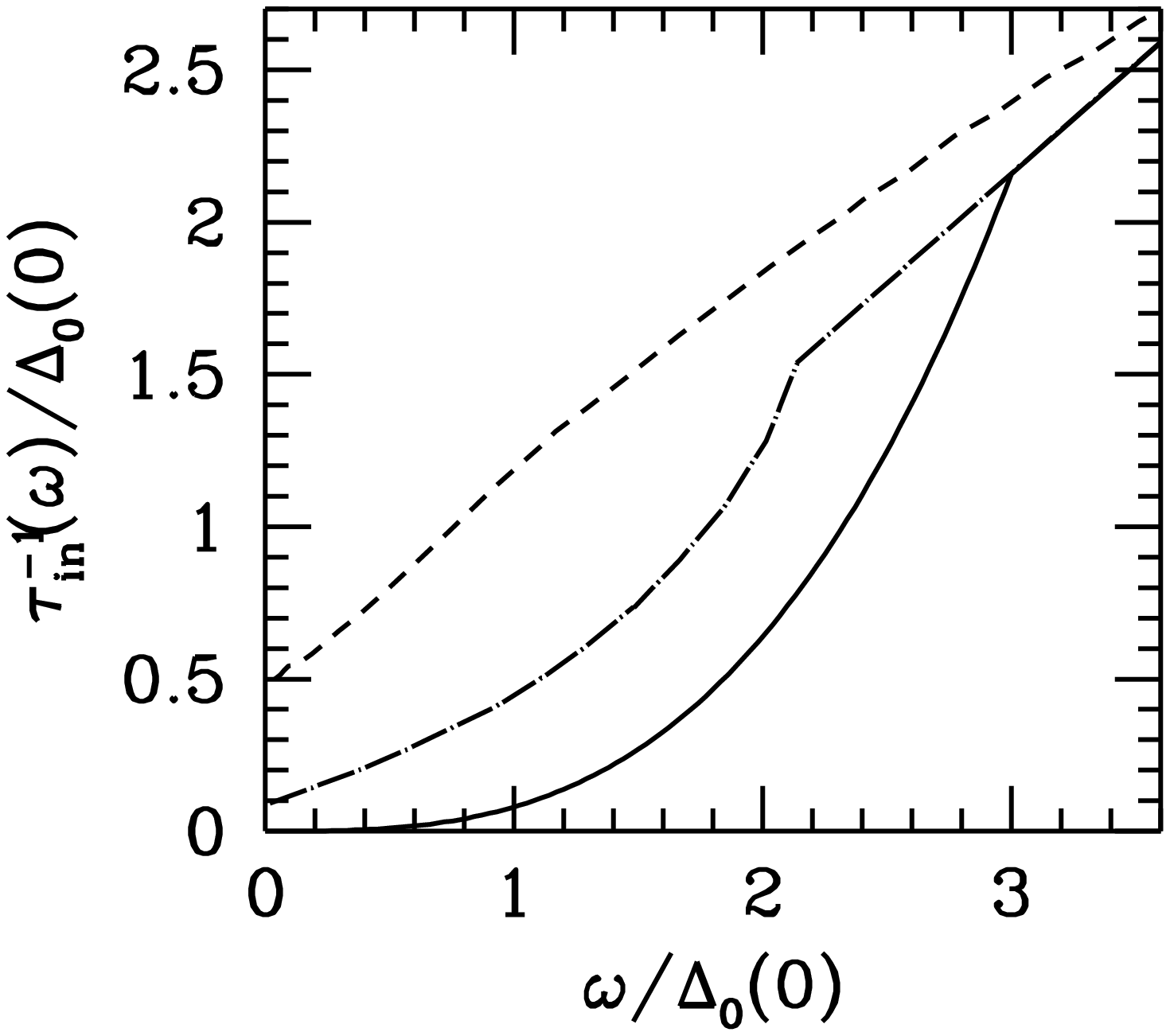,width=3.5in}
\vskip .2cm
\caption{Normalized intraplane inelastic relaxation rate
$\tau_{in}^{-1}(\omega)/\Delta_0(0)$
vs.\ $\omega/\Delta_0.$ Solid line: $T=0$;
dashed-dotted line: $T=0.8T_c$;
dashed line: $T=T_c$.  In all cases $\Delta_0(0)/T_c=3$.}
\end{figure}
\begin{figure}[p]
\leavevmode\centering\psfig{file=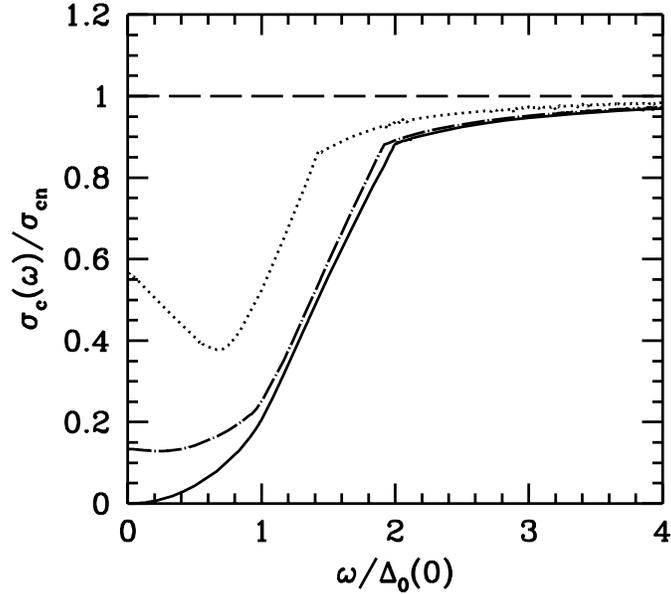,width=3.5in}
\vskip .2cm
\caption{Normalized $c$-axis conductivity  $\sigma_c(\omega) /\sigma_{cn}$
vs.\ $\omega/\Delta_0(0)$. Solid line: $T=0$; dashed-dotted line: $T=0.5T_c$;
dotted line: $T=0.8T_c$;
dashed line: $T=T_c$. Scattering anisotropy parameter $|V'/ V|^2$=0,
planar impurity scattering rate $\Gamma =0$, $\Delta_0(0)/T_c=3$.}
\end{figure}
\begin{figure}[p]
\leavevmode\centering\psfig{file=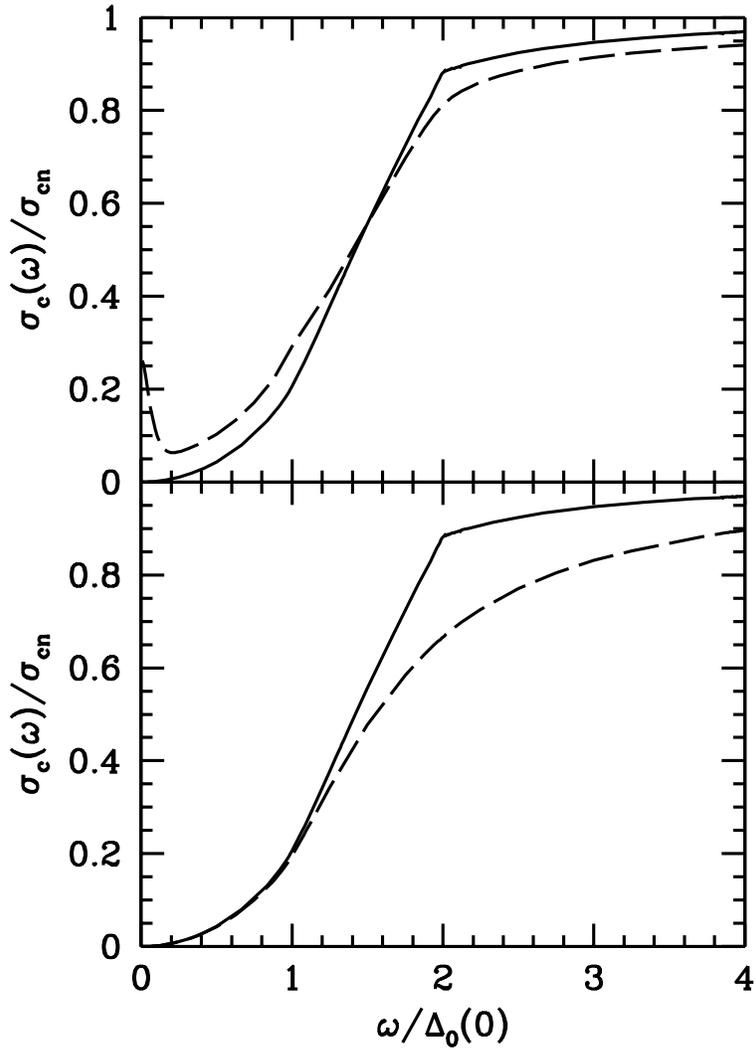,width=4.in}
\vskip .2cm
\caption{Normalized $c$-axis conductivity  $\sigma_c(\omega)/\sigma_{cn}$
vs.\ $\omega/\Delta_0(0)$ for $T=0$. a) $|{V'\over V}|^2=0$ with
scattering (dashed line, $\Gamma=0.1T_c$ and inelastic scattering
included) and without scattering (solid line, $1/\tau=0$); b) $1/\tau=0$
with $|{V'\over V}|^2=1$ (dashed line) and $|{V'\over V}|^2=0$ (solid
line).}
\end{figure}
\begin{figure}[p]
\leavevmode\centering\psfig{file=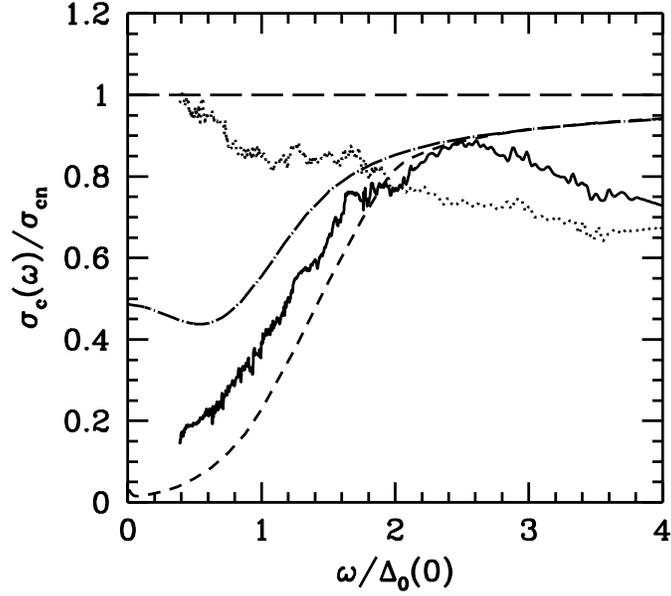,width=3.5in}
\vskip .2cm
\caption{Normalized $c$-axis conductivity  $\sigma_c(\omega) /\sigma_{cn}$
vs.\ $\omega/\Delta_0(0)$, $|V'/ V|^2=0,~\Gamma=0.02T_c,$
inelastic scattering included, $T=T_c$ (dashed line),
$T=0.8T_c$ (dashed-dotted line), and $T=0.1T_c$ (solid line), for
$\Delta_0/T_c=3$.
Data from Homes {\it et al.}\protect\cite{Homes} on $YBa_2 Cu_3 O_{6.95}$,
solid line: T=10K,
dashed line: T=100K.  $\sigma_{cn}$ normalized to 100K
data at 100 ${\rm cm^{-1}}$.
}
\end{figure}
\begin{figure}[p]
\leavevmode\centering\psfig{file=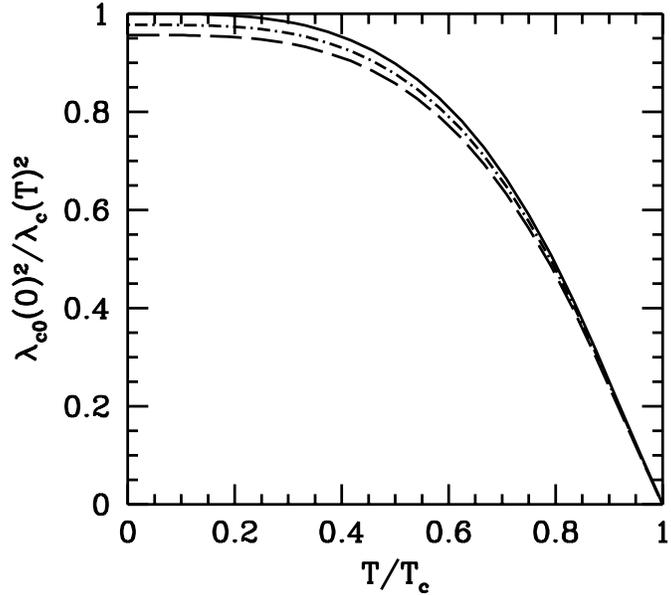,width=3.5in}
\vskip .2cm
\caption{Inverse $c$-axis London penetration depth $\lambda_c
(0)^2/\lambda_c(T)^2$
vs.\ normalized temperature $T/T_c$.  Solid line: intraplane impurity
scattering
rate $\Gamma/T_c=0$; dashed-dotted line:  $\Gamma/T_c=0.05$; dashed line:
$\Gamma/T_c=0.1$.  Inelastic scattering included.}
\end{figure}
\begin{figure}[p]
\leavevmode\centering\psfig{file=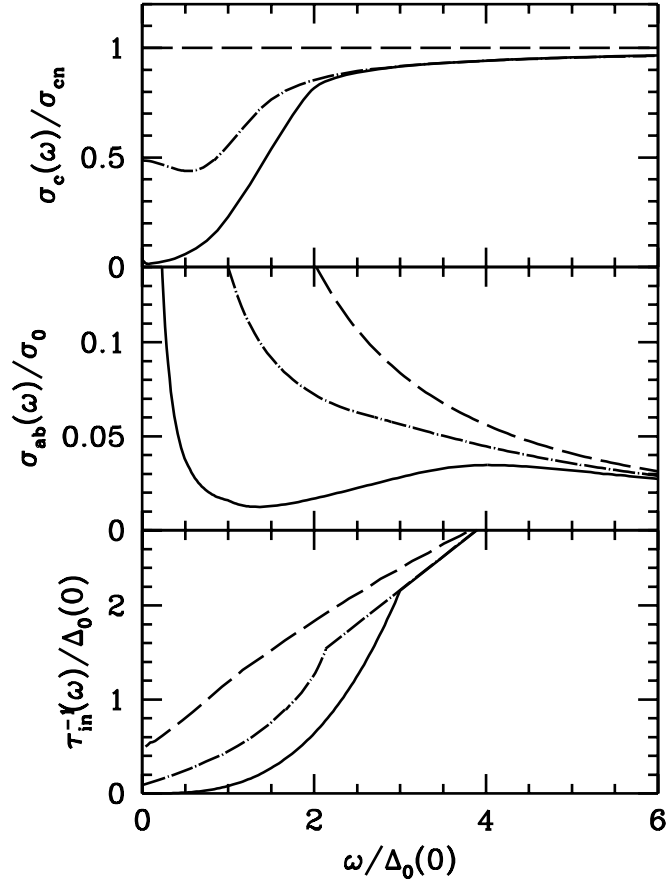,width=3.5in}
\vskip .2cm
\caption{
a) Normalized $c$-axis conductivity  $\sigma_c(\omega) /\sigma_{cn}$
vs.\ $\omega/\Delta_0(0)$; b)
$ab$-plane conductivity $\sigma_{ab}(\omega) /\sigma_0$
vs.\ $\omega/\Delta_0(0)$
(Note $\sigma_0\equiv\sigma_{ab}(\omega=0)$ at $T=T_c$);
c) quasiparticle relaxation rate $\tau_{in}^{-1}(\omega)/\Delta_0$
vs.\ $\omega/\Delta_0(0)$.  All panels show results for $\Gamma/T_c=0.02$,
$T=0.1T_c$ (solid line), $0.8T_c$ (dashed-dotted line),
and $T_c$ (dashed line).}
\end{figure}
\begin{figure}[p]
\leavevmode\centering\psfig{file=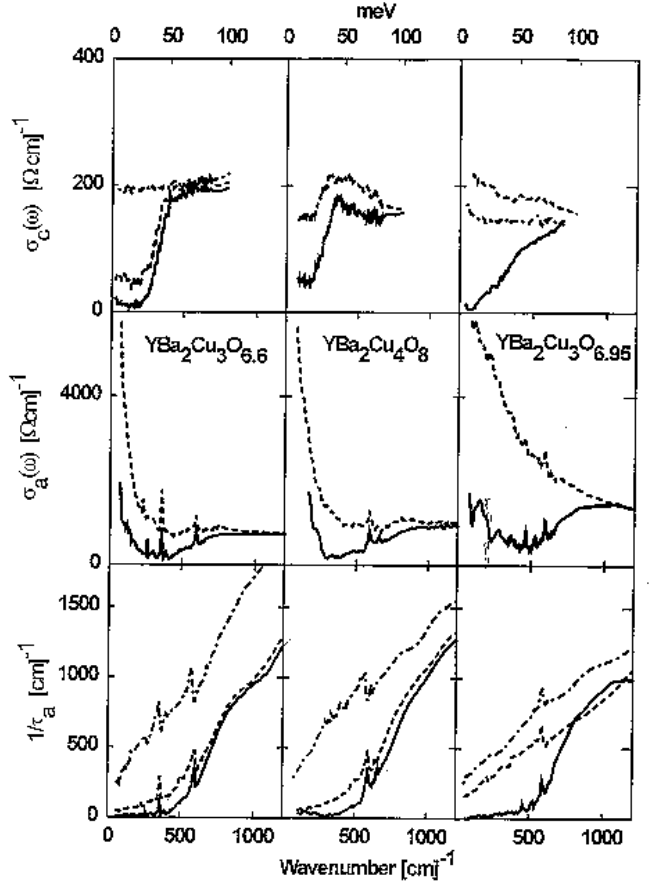,width=3.5in}
\vskip .2cm
\caption{Data of Basov {\it et al.}\protect\cite{Basov} for
$YBa_2 Cu_3 O_{6.95}$,
$YBa_2Cu_4 O_8$,
and $YBa_2 Cu_3 O_{6.6}$, in columns from left to right.  Dashed-dotted lines:
300K, dashed lines: $T\simeq T_c$, solid lines: 10K.  The $c$-axis conductivity
of single crystals is taken from Homes {\it et al.}\ and multiplied by a factor
of 4.  Lines indicate position of shoulder ($2\Delta_0$) in $\sigma_c$, and
twice this energy.  (Lines not included in LANL archived preprint.)}
\end{figure}
\begin{figure}[p]
\leavevmode\centering\psfig{file=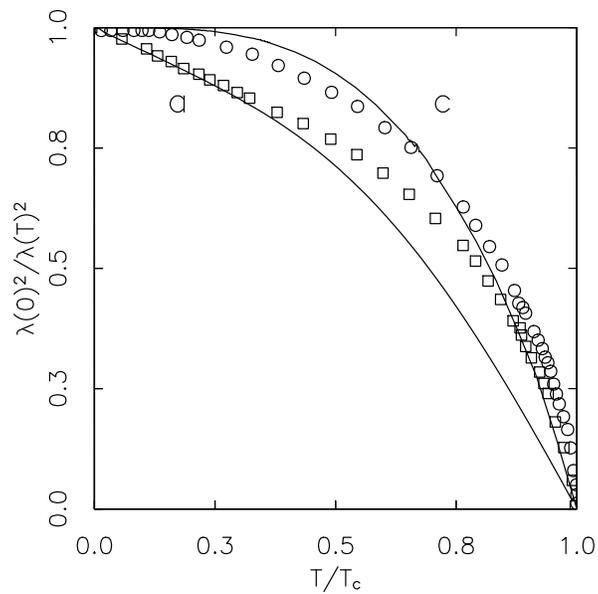,width=1.8in}
\vskip .2cm
\caption{Inverse $c$-axis and $ab$-plane London penetration depth
$\lambda_{c,ab} (0)^2/\lambda_{c,ab}(T)^2$
vs.\ normalized temperature $T/T_c$ for intraplane impurity scattering
rate $\Gamma=0$, no inelastic scattering included.  Data from Hardy
{\it et al.}\protect\cite{Hardyetal}
on $YBa_2 Cu_3 O_{6.95}$.}
\end{figure}

\end{document}